\documentclass[preprint,prd,aps,showpacs,showkeys,nofootinbib]{revtex4}
\usepackage{graphicx}
\usepackage{dcolumn}
\usepackage{bm}
\usepackage{color}
\usepackage[dvipsnames]{xcolor}
\usepackage{amssymb}
\usepackage{epstopdf}

\definecolor{black-blue}{RGB}{77,116,175}
\definecolor{black-yellow}{RGB}{231,162,33}
\definecolor{black-green}{RGB}{144,180,58}
\definecolor{black-red}{RGB}{246,95,50}

\begin{document}

\title{Lepton flavor violating decays $l_j \rightarrow l_i\gamma$ in the $U(1)_X$VLFM}
\author{Shuang Di$^{1,2,3}$, Wei-Hang Zhang$^{1,2,3}$, Zi-Xuan Su$^{1,2,3}$, Guo-Zhu Ning$^{1,2,3}$\footnote{ninggz@hbu.edu.cn}, Xing-Xing Dong$^{1,2,3,4}$\footnote{dongxx@hbu.edu.cn}, Shu-Min Zhao$^{1,2,3}$\footnote{zhaosm@hbu.edu.cn}}

\affiliation{$^1$ Department of Physics, Hebei University, Baoding 071002, China}
\affiliation{$^2$ Hebei Key Laboratory of High-precision Computation and Application of Quantum Field Theory, Baoding, 071002, China}
\affiliation{$^3$ Hebei Research Center of the Basic Discipline for Computational Physics, Baoding, 071002, China}
\affiliation{$^4$ Departamento de Fisica and CFTP, Instituto Superior T$\acute{e}$cnico, Universidade de Lisboa,
Av.Rovisco Pais 1,1049-001 Lisboa, Portugal}
\date{\today}

\begin{abstract}
In the $U(1)_X$ vector-like fermion model ($U(1)_X$VLFM), the Standard Model(SM) gauge group is extended with an additional $U(1)_X$ symmetry. Vector-like fermions and right-handed neutrinos are introduced, providing new sources of lepton flavor violation(LFV). In this paper, we perform a detailed study of the LFV decays $l_j \to l_i \gamma$ (with $j = \tau, \mu$; $i = \mu, e$; $i \neq j$) in this model. The numerical results show that, in certain parameter regions, the branching ratios of these processes can become large enough to be probed in future experiments. This work provides important theoretical guidance and constraints for exploring new physics beyond the SM.
\end{abstract}


\maketitle

\section{Introduction}
Neutrino oscillation experiments have demonstrated that neutrinos possess non-zero masses and revealed significant lepton flavor mixing\cite{n1,n2,n3,n4,n5}, directly challenging the fundamental assumption of lepton flavor conservation in the SM. Although the SM has been highly successful in describing elementary particle interactions, certain phenomena remain difficult to explain. For example, the Glashow-Iliopoulos-Maiani (GIM) mechanism leads to extremely suppressed branching ratios for LFV decays\cite{gim}. The SM prediction for the $\mu \rightarrow e\gamma$ branching ratio is as low as about $10^{-55}$, far below the current experimental sensitivity at 90\% confidence level\cite{551,553}. The experimental upper limits for the LFV processes $l_j \rightarrow l_i \gamma$ are as follows\cite{PDG,MEG1,BBr}
\begin{eqnarray}
&\mathrm{Br}(\mu \to e\gamma) < 4.2 \times 10^{-13}, \\
&\mathrm{Br}(\tau \to e\gamma) < 3.3 \times 10^{-8}, \\
&\mathrm{Br}(\tau \to \mu\gamma) < 4.2 \times 10^{-8}.
\end{eqnarray}
Therefore, if such decays are observed in future experiments, they would provide conclusive evidence for new physics beyond the SM.

To address the shortcomings of the SM in terms of neutrino masses, the fermion mass hierarchy, dark matter and so on, various extended frameworks have been proposed, such as the two-Higgs-doublet model\cite{HDM1,HDM2,HDM3}, the minimal supersymmetric Standard Model, and its extensions\cite{min}. In recent years, models featuring both vector-like fermions and an extra $U(1)$ gauge symmetry have attracted attention due to their rich phenomenological potential\cite{VLF1,VLF2,VLF3}. The $U(1)_X$VLFM adopted in this work extends the SM gauge group to $SU(3)_C \otimes SU(2)_L \otimes U(1)_Y \otimes U(1)_X$. This model introduces three generations of right-handed neutrinos $\nu_R$, two scalar singlets $\phi$ and $S$, as well as one generation of vector-like quarks, vector-like leptons, and vector-like neutrino. The CP-even components of the neutral scalar fields mix, forming a $3\times3$ mass matrix, while the light neutrino masses are generated at tree level via the seesaw mechanism\cite{ss}. All SM particles carry zero $U(1)_X$ charges, ensuring the theory is anomaly-free\cite{free1,free2}. These new fields introduce additional sources of flavor mixing in the lepton sector: the mixing between vector-like leptons and SM leptons gives rise to contributions at the one-loop level for the $l_j \to l_i \gamma$ processes that differ from those of the SM. These new contents provide additional one-loop contributions to LFV processes and can significantly enhance the branching ratios of decays such as $l_j \to l_i \gamma$\cite{VLF1,smvlf2}.

This paper studies the LFV decays $l_j \to l_i \gamma$ within the framework of the $U(1)_X$VLFM, where $j = \tau, \mu$, $i = \mu, e$, and $i \neq j$. The new particles and interactions introduced in this model, including the mixing between SM fermions and vector-like fermions, the extra gauge boson $Z'$, and the extended scalar fields, can all contribute to the $l_j \to l_i \gamma$ processes at one-loop level\cite{Zp,Wp}. Compared with the SM, the $U(1)_X$VLFM provides additional sources of LFV, potentially enhancing the branching ratios to levels detectable by current or upcoming experiments.

The structure of this paper is as follows. Sec.II briefly summarizes the basic content of the $U(1)_X$VLFM, including the gauge group, particle spectrum, mass mixing, and the interaction vertices in the relevant Lagrangian. Sec.III derives the analytic expression for the branching ratio of $l_j \to l_i \gamma$ in this framework, presenting the explicit forms of the Wilson coefficients and the definitions of the one-loop integral functions. Sec.IV performs parameter scans and plotting, along with numerical analysis. Sec.V presents the conclusions.
\section{THE $U(1)_X$VLFM}
In the $U(1)_X$VLFM, besides the SM leptons, one generation of vector-like leptons is introduced. These vector-like leptons mix with the SM leptons via the vacuum expectation values of the scalar fields $\phi$ and $S$, leading to a non-diagonal structure in the mass matrix\cite{smlmix1,smlmix2}. This mixing provides new contributions to LFV processes, in particular to the radiative decays $l_j \to l_i \gamma$\cite{radLFV1,radLFV2}. The new fields beyond the SM are listed in Table~\ref{tab}. In the following, we describe in detail the Lagrangian, mass mixing, and effective interactions relevant to those processes in the model.

\begin{table}[ht]
\centering
\caption{Properties of new particles introduced in the model}
\begin{tabular}{|c|c|c|c|c|}
\hline
Field & $SU(3)_C$ & $SU(2)_L$ & $U(1)_Y$ & $U(1)_X$ \\
\hline
$\phi$ & 1 & 1 & 0 & $Q_a + Q_b$ \\
\hline
$S$ & 1 & 1 & 0 & $Q_a$ \\
\hline
$\nu_R$ & 1 & 1 & 0 & 0 \\
\hline
$d_{XL}$ & 3 & 1 & -1/3 & $Q_a$ \\
\hline
$u_{XL}$ & 3 & 1 & 2/3 & $-Q_a$ \\
\hline
${\bar{d}_{XR}}$ & $\bar{3}$ & 1 & 1/3 & $Q_b$ \\
\hline
${\bar{u}_{XR}}$ & $\bar{3}$ & 1 & -2/3 & $-Q_b$ \\
\hline
$e_{XL}$ & 1 & 1 & -1 & $Q_a$ \\
\hline
$\nu_{XL}$ & 1 & 1 & 0 & $-Q_a$ \\
\hline
${\bar{e}_{XR}}$ & 1 & 1 & 1 & $Q_b$ \\
\hline
${\bar{\nu}_{XR}}$ & 1 & 1 & 0 & $-Q_b$ \\
\hline
\end{tabular}
\label{tab}
\end{table}

In the $U(1)_X$VLFM, the scalar sector consists of one $SU(2)_L$ Higgs doublet $H$ and two Higgs singlets $\phi$ and $S$. Their explicit expressions are given by
{\begin{eqnarray}
&&{H=\left(\begin{array}{c}H^{+} \\ H^{0}\end{array}\right)\;}, ~~~~~H^0=\frac{1}{\sqrt{2}}(v+\phi_H+i\sigma_H),
\nonumber\\&&\hspace{0cm}\phi=\frac{1}{\sqrt{2}}(v_P+\phi_P+i\sigma_P),~~~S=\frac{1}{\sqrt{2}}(v_S+\phi_S+i\sigma_S).
\end{eqnarray}}
where $v$, $v_P$ and $v_S$ denote the nonzero vacuum expectation values (VEVs) of the Higgs fields $H$, $\phi$ and $S$.

In the $U(1)_X$VLFM, the Lagrangian relevant to the scalar potential and Yukawa interactions can be written as
\begin{eqnarray}
\mathcal{L} \supset & -\mu_H^2 H^\dagger H - \mu_P^2 |\phi|^2 - \mu_X^2 |S|^2+ \lambda_H (H^\dagger H)^2 + \lambda_P |\phi|^4 + \lambda_X |S|^4
\nonumber\\&&\hspace{-11cm} + \lambda_{HP}(H^\dagger H)|\phi|^2 + \lambda_{HX}(H^\dagger H)|S|^2 + \lambda_{PX}|S|^2|\phi|^2
\nonumber\\&&\hspace{-11cm} - S\, {\bar{d}_{XL,k}} Y_{XD,jk}^* d_{R,j} - S\, {\bar{u}_{R,j}} Y_{XU,jk} u_{XL,k} - S\, {\bar{e}_{XL,k}} Y_{XE,jk}^* e_{R,j}
\nonumber\\&&\hspace{-11cm} - S\, {\bar{\nu}_{R,j}} Y_{XN,jk} \nu_{XL,k} + \text{h.c.}
\nonumber\\&&\hspace{-11cm} - \phi\, {\bar{d}_{XL,k}} Y_{PD,jk}^* d_{XR,j} - \phi\, {\bar{u}_{XR,j}} Y_{PU,jk} u_{XL,k} - \phi\, {\bar{e}_{XL,k}} Y_{PE,jk}^* e_{XR,j}
\nonumber\\&&\hspace{-11cm} - \phi\, {\bar{\nu}_{XR,j}} Y_{PN,jk} \nu_{XL,k} + \text{h.c.}
\nonumber\\&&\hspace{-11cm} - Y_{u,jk}^* \bar{q}_{L,k} {\tilde{H}} u_{R,j} + Y_{d,jk}^* \bar{q}_{L,k} {H} d_{R,j} + Y_{e,jk}^* \bar{l}_k \tilde{H} e_{R,j} + \text{h.c.} .
\end{eqnarray}

In the $U(1)_X$VLFM, the particles beyond the SM and their quantum number assignments are summarized in Table~\ref{tab}. According to Ref.\cite{wx}, the gauge sector of the SM is itself anomaly-free. In the present extended framework, the introduction of one generation of vector-like fermions is precisely to ensure the overall anomaly cancellation. Concretely, the model satisfies the following anomaly cancellation conditions
\begin{enumerate}
  \item The pure $SU(3)_C$ and pure $SU(2)_L$ anomalies are the same as those in the SM and strictly vanish.
  \item Mixed anomalies involving one $SU(3)_C$ or one $SU(2)_L$ gauge boson are proportional to $Tr[t^a] = 0$ or $Tr[\tau^a] = 0$, hence they also vanish automatically.
  \item Mixed anomalies involving one $U(1)_Y$ (or $U(1)_X$) and two $SU(3)_C$ bosons are proportional to $\sum_q Y_q^Y$ and $\sum_q Y_q^X$, respectively.
  \item Similarly, mixed anomalies involving one $U(1)_Y$ (or $U(1)_X$) and two $SU(2)_L$ bosons are proportional to $\sum_L Y_L^Y$ and $\sum_L Y_L^X$, respectively.
  \item The pure $U(1)$ anomalies consist of the following four types
        \begin{eqnarray}
        Tr[Y^Y Y^Y Y^Y] = \sum_n (Y_n^Y)^3,\qquad
        Tr[Y^X Y^X Y^X] = \sum_n (Y_n^X)^3,
       \nonumber\\&&\hspace{-11cm} Tr[Y^X Y^Y Y^Y] = \sum_n Y_n^X (Y_n^Y)^2,\qquad
        Tr[Y^Y Y^X Y^X] = \sum_n Y_n^Y (Y_n^X)^2.
        \end{eqnarray}
  \item The mixed gravitational anomalies involving one $U(1)$ gauge boson are proportional to $\sum_n Y_n^Y$ or $\sum_n Y_n^X$.
\end{enumerate}

By summing over the quantum numbers given in Table~I, one can verify that all the above anomaly conditions are satisfied. Therefore, the
$U(1)_X$VLFM is a theoretically consistent and anomaly-free model.

In the $U(1)_X$VLFM, the anomaly structures that do not involve the $U(1)_X$ gauge group are identical to those of the SM and thus automatically satisfy the anomaly-free condition. For the anomaly terms introduced by the additional $U(1)_X$ symmetry, their complete cancellation is achieved by the introduction of one generation of vector-like fermions. Consequently, all gauge and mixed gravitational anomalies are successfully eliminated, ensuring the theoretical consistency of the model. Furthermore, the coexistence of two Abelian gauge groups, $U(1)_Y$ and $U(1)_X$, gives rise to a novel effect not present in the SM: gauge kinetic mixing. Even if this mixing is set to zero at the $M_{\text{GUT}}$ scale, it can still be radiatively generated through the renormalization group equations.

In this model, the covariant derivative is
{\begin{eqnarray}
&&D_\mu=\partial_\mu-i\left(\begin{array}{cc}Y^Y,&Y^X\end{array}\right)
\left(\begin{array}{cc}g_{Y},&g{'}_{{YX}}\\g{'}_{{XY}},&g{'}_{{X}}\end{array}\right)
\left(\begin{array}{c}A_{\mu}^{\prime Y} \\ A_{\mu}^{\prime X}\end{array}\right)\;,
\end{eqnarray}}
where $A^{'Y}_\mu$ and $A^{'X}_\mu$ denote the gauge fields of $U(1)_Y$ and $U(1)_X$. A basis rotation can be performed using an orthogonal matrix $R$ ($R^T R = 1$). The transformed form is given by
{\begin{eqnarray}
&&\left(\begin{array}{cc}g_{Y},&g{'}_{{YX}}\\g{'}_{{XY}},&g{'}_{{X}}\end{array}\right)R^T=
\left(\begin{array}{cc}g_{1},&g_{{YX}}\\0,&g_{{X}}\end{array}\right),
\end{eqnarray}}
which redefines the $U(1)$ gauge fields as
{\begin{eqnarray}
&&R\left(\begin{array}{c}A_{\mu}^{\prime Y} \\ A_{\mu}^{\prime X}\end{array}\right)=\left(\begin{array}{c}A_{\mu}^{ Y} \\ A_{\mu}^{ X}\end{array}\right).
\end{eqnarray}}

Here $g_X$ is the gauge coupling constant of the $U(1)_X$ symmetry, while $g_{YX}$ describes the gauge kinetic mixing between the $U(1)_Y$ and $U(1)_X$ gauge groups. At tree level, the neutral gauge bosons $A_\mu^Y$, $V_\mu^3$, and $A_\mu^X$ undergo mixing, resulting in the mass matrix in the $(A_\mu^Y, V_\mu^3, A_\mu^X)$ basis

{\begin{eqnarray}
\left( \begin{array}{ccc}
\frac{1}{4}g_{1}^2v^2 & -\frac{1}{4}g_{1}g_{2}v^2 & \frac{1}{4}g_{1}g_{YX}v^2 \\
-\frac{1}{4}g_{1}g_{2}v^2 & \frac{1}{4}g_{2}^2v^2 &-\frac{1}{4}g_{2}g_{YX}v^2 \\
\frac{1}{4}g_{1}g_{YX}v^2 & -\frac{1}{4}g_{2}g_{YX}v^2 & \frac{1}{4}g_{YX}^2v^2+\frac{1}{4}g_{X}^2\xi^2
\end{array} \right),
\label{eq1}
\end{eqnarray}}
with $\xi^2 = 4(Q_a + Q_b)^2 v_P^2 + 4Q_a^2 v_S^2$. The diagonalization of the above matrix can be realized by a rotation operation involving the weak mixing angle $\theta_W$ and the additional angle $\theta_W'$.
{\begin{eqnarray}
\left(
\begin{array}{c}
\gamma_\mu \\
Z_\mu \\
Z'_\mu
\end{array}
\right)
=
\left(
\begin{array}{ccc}
\cos\theta_W & \sin\theta_W & 0 \\
-\sin\theta_W \cos\theta'_W & \cos\theta_W \cos\theta'_W & \sin\theta'_W \\
\sin\theta_W \sin\theta'_W & -\cos\theta_W \sin\theta'_W & \cos\theta'_W
\end{array}
\right)
\left(
\begin{array}{c}
A^Y_\mu \\
V^3_\mu \\
A^X_\mu
\end{array}
\right).
\end{eqnarray}}
The additional mixing angle $\theta'_W$, which governs the couplings of the $Z$ and $Z'$ bosons, is given by
\begin{equation}
\sin^2 \theta'_W = \frac{1}{2} - \frac{(g_{YX}^2 - g_1^2 - g_2^2)v^2 + g_X^2 \xi^2}{2\sqrt{(g_{YX}^2 + g_1^2 + g_2^2)^2 v^4 + 2g_X^2(g_{YX}^2 - g_1^2 - g_2^2)v^2 \xi^2 + g_X^4 \xi^4}}.
\end{equation}

The exact mass eigenvalues from Eq.~(\ref{eq1}) are
\begin{eqnarray}
&&\hspace{0cm}m_\gamma^2 = 0,
\nonumber\\&&\hspace{0cm}m_{Z,Z'}^2 = \frac{1}{8}\left((g_1^2 + g_2^2 + g_{YX}^2)v^2 + g_X^2 \xi^2\right.
\nonumber\\&&\hspace{0cm}\quad \left.\mp \sqrt{\left[(g_1^2 + g_2^2 + g_{YX}^2)v^2 + g_X^2 \xi^2\right]^2 - 4(g_1^2 + g_2^2)g_X^2 v^2 \xi^2}
\right).\label{zp}
\end{eqnarray}

The VEVs of the Higgs fields satisfy the following system of equations
\begin{eqnarray}
2\lambda_H v^2 - 2\mu_H^2 + \lambda_{HP} v_P^2 + \lambda_{HX} v_S^2 &= 0, \\
2\lambda_X v_S^2 - 2\mu_X^2 + \lambda_{HX} v^2 + \lambda_{PX} v_P^2 &= 0, \\
2\lambda_P v_P^2 - 2\mu_P^2 + \lambda_{HP} v^2 + \lambda_{PX} v_S^2 &= 0.
\end{eqnarray}

In the basis $(e_L, e_{XL})$ and {$(\bar{e}_R, \bar{e}_{XR})$}, the mass matrix of leptons is given by
{\begin{eqnarray}
&&m_e=\left(\begin{array}{cc} \frac{1}{\sqrt{2}} v Y_e^T,&0\\ \frac{1}{\sqrt{2}} {v_S} Y_{XE}^T,&\frac{1}{\sqrt{2}} v_P Y_{PE}^T\end{array}\right),
\end{eqnarray}}
This matrix is diagonalized by $U_L^e$ and $U_R^e$
\begin{eqnarray}
&&U_L^{e,*} \, m_e \, U_R^{e,\dagger} = m_e^{\text{dia}},
\end{eqnarray}
with
\begin{eqnarray}
&&\hspace{0cm}e_{L,i}= \sum_{t_2} U_{L,ji}^{e,*} E_{L,j},
~~~e_{XL,i}= \sum_{t_2} U_{L,ji}^{e,*} E_{L,j},
\nonumber\\&&\hspace{0cm}e_{R,i}= \sum_{t_2} U_{R,ij}^{e} E_{R,j}^*,
~~~e_{XR,i}= \sum_{t_2} U_{R,ij}^{e} E_{R,j}^*.
\end{eqnarray}

In the basis $(\nu_L, {\bar{\nu}_R}, \nu_{XL}, {\bar{\nu}_{XR})}$, the mass matrix of neutrinos is given by
\begin{eqnarray}
m_\nu = \left(\begin{array}{cccc}
0 & \frac{1}{\sqrt{2}} v Y_\nu^T & 0 & 0 \\[1em]
\frac{1}{\sqrt{2}} v Y_\nu & 0 & \frac{1}{\sqrt{2}} v_S Y_{XN}^T & 0 \\[1em]
0 & \frac{1}{\sqrt{2}} v_S Y_{XN} & 0 & \frac{1}{\sqrt{2}} v_P Y_{PN}^T \\[1em]
0 & 0 & \frac{1}{\sqrt{2}} v_P Y_{PN} & 0
\end{array}\right),
\end{eqnarray}
where $v$, $v_S$, $v_P$ are the VEVs of $H$, $S$, and $\phi$, respectively. This matrix is diagonalized by a unitary matrix $U^\nu$
\begin{eqnarray}
U^{\nu,*} \, m_\nu \, U^{\nu,\dagger} = m_\nu^{\text{dia}},
\end{eqnarray}
with
\begin{eqnarray}
\nu_{L,i} &=& \sum_j U_{ji}^{\nu,*} \nu_{L,j}, \qquad
\nu_{R,i}^* = \sum_j U_{ji}^{\nu} \nu_{R,j}^{*}, \\
\nu_{XL,i} &=& \sum_j U_{j,i}^{\nu,*} \nu_{L,j}, \qquad
\nu_{XR,i}^* = \sum_j U_{j,i}^{\nu} \nu_{R,j}^{*}.
\end{eqnarray}

{In the basis $(\phi_H,\phi_S,\phi_P)$, the mass matrix of Higgs is given by
{\begin{eqnarray}
&&m_h^2=\left(\begin{array}{ccc} m_{\phi_H\phi_H} & -\lambda_{HX} v v_S & -\lambda_{HP} v v_P\\
-\lambda_{HX} v v_S & m_{\phi_S\phi_S} & -\lambda_{PX} v_P v_S\\
-\lambda_{HP} v v_P & -\lambda_{PX} v_P v_S & m_{\phi_P\phi_P}
\end{array}\right),
\end{eqnarray}}
with
\begin{eqnarray}
&&m_{\phi_H\phi_H} = \frac{1}{2}\big(-6\lambda_H v^2 - \lambda_{HP} v_P^2 - \lambda_{HX} v_S^2\big) + \mu_H^2,
\\
&&m_{\phi_S\phi_S} = \frac{1}{2}\big(-6\lambda_X v_S^2 - \lambda_{HX} v^2 - \lambda_{PX} v_P^2\big) + \mu_X^2,
\\
&&m_{\phi_P\phi_P} = \frac{1}{2}\big(-6\lambda_P v_P^2 - \lambda_{HP} v^2 - \lambda_{PX} v_S^2\big) + \mu_P^2.
\end{eqnarray}
This matrix is diagonalized by $Z^H$
\begin{eqnarray}
&&Z^{H} \, m_h^2 \, Z^{H,\dagger} = m_{2,h}^{\text{dia}},
\end{eqnarray}
with
\begin{eqnarray}
&&\hspace{0cm}\phi_{H}= \sum_{j} Z_{j1}^{H} h_{j},
~~~\phi_{S}= \sum_{j} Z_{j2}^{H} h_{j},
~~~\hspace{0cm}\phi_{P}= \sum_{j} Z_{j3}^{H} h_{j}.
\end{eqnarray}}
\section{ANALYTICAL FORMULA}
\begin{figure}[ht]
\setlength{\unitlength}{5mm}
\centering
\includegraphics[width=3.5in]{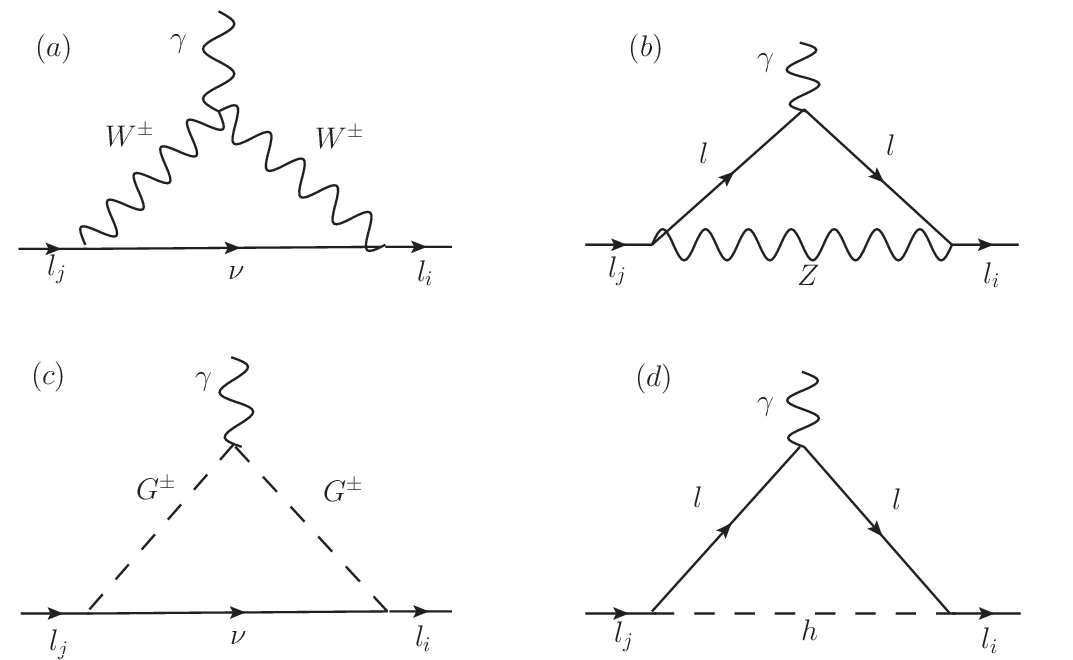}
\setlength{\unitlength}{5mm}
\caption{One-loop diagrams for $l_j\rightarrow l_i\gamma$.}
\label{t1}
\end{figure}

In this section, we study the LFV processes $l_j\rightarrow l_i\gamma$ $(j=\tau, \mu,~i=\mu, e$ and $i\neq j)$ in the $U(1)_X$VLFM. The relevant Feynman diagrams are shown in Fig.~\ref{t1}. Under the on-shell condition for the external leptons, the amplitude of the processes $l_j\rightarrow l_i\gamma$ can be written in the general form\cite{2015prd}
\begin{eqnarray}
\mathcal{M} = e \varepsilon^{\mu} \bar{u}_i (p + q) \left[ q^2 \gamma_{\mu} (C_1^L P_L + C_1^R P_R)
+ m_{l_j} i \sigma_{\mu \nu} q^{\nu} (C_2^L P_L + C_2^R P_R) \right] u_j(p),
\end{eqnarray}
where $p$ denotes the four-momentum of the incoming lepton, $q$ denotes the four-momentum of the photon, and $m_{l_j}$ represents the mass of the charged lepton in the $j$-th generation.
$u_i(p+q)$ and $u_j(p)$ correspond to the Dirac spinors of the external leptons.
The complete Wilson coefficients $C_1^L$, $C_1^R$, $C_2^L$, and $C_2^R$ are derived from the coherent sum of the scattering amplitudes of these diagrams.

Because the three light neutrinos mix with the three heavy neutrinos and vector like neutrinos, the virtual $W$-boson loop diagrams Fig.~\ref{t1} (a) contribute corrections to the LFV processes $l_j \to l_i \gamma$.
The corresponding Wilson coefficients, denoted by $C_\alpha^{L,R}(W)$ ($\alpha=1,2$), are given by
\begin{eqnarray}
&&\hspace{0cm}C_1^L(W)= \sum_{F=\nu} \frac{-1}{2 m_W^2} H_L^{W^{\pm}F\bar{l}_i} H_L^{W^{\pm} l_j \bar{F}} \left[ I_2(x_F, x_W) + I_1(x_F, x_W) \right],
\nonumber\\&&\hspace{0cm}C_2^L(W)= \sum_{F=\nu} \frac{1}{m_W^2} H_L^{W^{\pm}F\bar{l}_i} H_L^{W^{\pm} l_j \bar{F}} \left( 1 + \frac{m_{l_i}}{m_{l_j}} \right) \left[ 2 I_2(x_F, x_W) - \frac{1}{3} I_1(x_F, x_W) \right],
\nonumber\\&&\hspace{0cm}C_\alpha^R(W)= 0, \quad \alpha=1,2.\label{C1}
\end{eqnarray}
Here, $x_F = m_F^2/m_W^2$ and $x_W = m_W^2/m_W^2 = 1$, with $m_F$ being the mass of the neutrino eigenstate $F$. $H_{L;R}^{W^{\pm}F\bar{l}_i}$ and $H_{L;R}^{W^{\pm} l_j \bar{F}}$ are the corresponding couplings of the left(right)-handed parts in the Lagrangian. The specific forms are shown as follows
\begin{eqnarray}
&&\hspace{0cm}H_{L}^{W^{\pm} l_j \bar{F}}=-\frac{1}{\sqrt{2}} g_2 \sum_{a=1}^{3} U_{L,ja}^{e,*} U_{ia}^{V},\\
\nonumber\\&&\hspace{0cm}H_{L}^{W^{\pm}F\bar{l}_i}=- \frac{1}{\sqrt{2}} g_2 \sum_{a=1}^{3} U_{ja}^{V,*} U_{L,ia}^{e}.
\end{eqnarray}

The one-loop functions $I_i(x_1, x_2)$, $i=1\ldots3$ are shown here
\begin{eqnarray}
&&\hspace{0cm}I_1(x_1,x_2) = \frac{1}{96\pi^2} \bigg[ \frac{11 + 6 \ln x_2}{(x_2 - x_1)} - \frac{15x_2 + 18x_2 \ln x_2}{(x_2 - x_1)^2} + \frac{6x_2^2 + 18x_2^2 \ln x_2}{(x_2 - x_1)^3}\nonumber\\&&\hspace{2cm} + \frac{6x_1^3 \ln x_1- 6x_2^3 \ln x_2}{(x_2 - x_1)^4}\bigg],\\
\nonumber\\&&\hspace{0cm}I_2(x_1,x_2) = \frac{1}{32\pi^2} \left[ \frac{3 + 2 \ln x_2}{(x_2 - x_1)} - \frac{2x_2 + 4x_2 \ln x_2}{(x_2 - x_1)^2} - \frac{2x_1^2 \ln x_1}{(x_2 - x_1)^3} + \frac{2x_2^2 \ln x_2}{(x_2 - x_1)^3} \right],\\
\nonumber\\&&\hspace{0cm}I_3(x_1,x_2) = \frac{1}{16\pi^2} \left[ \frac{1 + \ln x_2}{(x_2 - x_1)} + \frac{x_1 \ln x_1 - x_2 \ln x_2}{(x_2 - x_1)^2} \right].
\end{eqnarray}
For the virtual $Z$-boson loop diagrams Fig.~\ref{t1}(b), the mixing between SM leptons and vector like leptons  contributes corrections to the LFV processes $l_j \to l_i \gamma$. The corresponding Wilson coefficients, denoted by $C_\alpha^{L,R}(Z)$ ($\alpha=1,2$), are given by

\begin{eqnarray}
&&\hspace{0cm}C_1^L(Z)= \sum_{F=l} \frac{1}{3 m_W^2} H_L^{ZF\bar{l}_i} H_L^{Z^* l_j \bar{F}} \left[ 3I_2(x_F, x_Z) + I_1(x_F, x_Z) \right],
\nonumber\\&&\hspace{0cm}C_2^L(Z)= \sum_{F=l} \frac{m_F}{m_{l_j}m_W^2} H_L^{ZF\bar{l}_i} H_L^{Z^* l_j \bar{F}} \left( 1 + \frac{m_{l_i}}{m_{l_j}} \right) \left[ I_2(x_F, x_Z)+2I_3(x_F, x_Z) \right],
\nonumber\\&&\hspace{0cm}C_a^R(Z) = C_a^L(Z)\big|_{L \leftrightarrow R}, \qquad a = 1,2.\label{C2}
\end{eqnarray}
where $H_{L;R}^{ZF\bar{l}_i}$ and $H_{L;R}^{Z^* l_j \bar{F}}$ are the corresponding couplings of the left(right)-hand parts in the Lagrangian.

The contributions from the one-loop triangle diagrams involving virtual neutrinos and {charged Goldstone bosons $G^\pm$} in Fig.~\ref{t1}(c) are denoted by $C_{a}^{L,R}(H)$ ($a=1,2$) and are expressed as follows
\begin{eqnarray}
&&\hspace{0cm}C_1^L(H) = \sum_{F=\nu}\frac{1}{6 m_W^2} H_R^{{G^{\pm}} F \bar{l}_i} H_L^{{G^{\pm}} l_j \bar{F}} \, I_1(x_F, x_S),
\nonumber\\&&\hspace{0cm}C_2^L(H) = \sum_{F=\nu} \frac{m_F}{m_{l_j} m_W^2} H_R^{{G^{\pm}} F \bar{l}_i} H_L^{{G^{\pm}} l_j \bar{F}} \, \bigl[ I_2(x_F, x_S) - I_3(x_F, x_S) \bigr],
\nonumber\\&&\hspace{0cm}C_a^R(H) = C_a^L(H)\big|_{L \leftrightarrow R}, \qquad a = 1,2.\label{C3}
\end{eqnarray}

{We adopt the Feynman-'t Hooft gauge with $\xi=1$. In this gauge, the charged Goldstone boson satisfies $m_{G^\pm}^2=\xi m_W^2=m_W^2$, so that $x_S=x_W$.} $H_{L,R}^{{G^{\pm}} F \bar{l}}$ denote the left- and right-handed couplings of the ${G^{\pm}}$-$F$-$\bar{l}$ vertex. The specific forms are shown as follows
\begin{eqnarray}
H_L^{{G^{\pm}} l_j \bar{F}}=- \sum_{a,~b=1}^{3} U_{L,jb}^{e,*} U_{i3+a}^{V,*} Y_{\nu,ab}, ~~~H_R^{{G^{\pm}} l_j \bar{F}}= \sum_{a,~b=1}^{3} Y_{e,ab}^* U_{R,ja}^{e} U_{ib}^{V}.
\end{eqnarray}

The contributions from the one-loop triangle diagrams with virtual leptons and neutral scalars  in Fig.~\ref{t1}(d) are denoted by $C_{\alpha}^{L,R}(h)$ ($\alpha = 1, 2$) and are expressed as follows

\begin{eqnarray}
&&\hspace{0cm}C_1^L(h) = \sum_{F=l} \sum_{S=h} \frac{1}{6m_W^2} H_R^{hF\bar{l}} H_L^{h^* l F} \bigl[ I_3(x_F, x_S) - 2I_4(x_F, x_S) - I_1(x_F, x_S) \bigr],
\nonumber\\&&\hspace{0cm}C_2^L(h) = \sum_{F=l} \sum_{S=h} \frac{m_F}{m_{l_j} m_W^2} H_L^{hF\bar{l}} H_L^{h^* l F} \bigl[ I_3(x_F, x_S) - I_4(x_F, x_S) - I_1(x_F, x_S) \bigr],
\nonumber\\&&\hspace{0cm}C_\alpha^R(h) = C_\alpha^L(h)\big|_{L \leftrightarrow R}, \qquad \alpha = 1, 2.\label{C4}
\end{eqnarray}

$H_{L,R}^{hF\bar{l}}$ denote the left- and right-handed couplings in the $h$-$F$-$\bar{l}$ interaction vertices. The specific forms are shown as follows
\begin{eqnarray}
\nonumber\\&&\hspace{0cm}H_{L}^{hFl}=-\frac{1}{\sqrt{2}}(  \sum_{a=1}^{3}U_{L,ja}^{e,*} U_{R,ia}^{e,*} Y_{e,a} Z_{k1}^{H}
+\sum_{a=1}^{3}U_{L,j4}^{e,*}U_{R,ia}^{e,*} Y_{XE,a} Z_{k2}^{H}
\nonumber\\&&\hspace{1.5cm}+U_{L,j4}^{e,*} U_{R,i4}^{e,*} Y_{PE} Z_{k3}^{H}),\\
\nonumber\\&&\hspace{0cm}H_{R}^{hFl}=- \frac{1}{\sqrt{2}} (
\sum_{a=1}^{3} Y_{e,a}^* U_{R,ja}^{e} U_{L,ia}^{e} Z_{k1}^{H}+ \sum_{a=1}^{3} Y_{XE,a}^* U_{R,ja}^{e} U_{L,i4}^{e} Z_{k2}^{H}
\nonumber\\&&\hspace{1.5cm}+ Y_{PE}^* U_{R,j4}^{e} U_{L,i4}^{e} Z_{k3}^{H}).
\label{hll}
\end{eqnarray}

The specific form of function $I_4(x_1, x_2)$
\begin{eqnarray}
I_4(x_1, x_2) = \frac{1}{16\pi^2} \left[ -\frac{1 + \ln x_1}{(x_2 - x_1)} - \frac{x_1 \ln x_1 - x_2 \ln x_2}{(x_2 - x_1)^2} \right].
\end{eqnarray}

The total coefficients are the sum of Eqs. (\ref{C1}), (\ref{C2}), (\ref{C3}) and (\ref{C4})

\begin{eqnarray}
C_a^{L,R} = C_a^{L,R}(H) + C_a^{L,R}(h) + C_a^{L,R}(W) +C_a^{L,R}(Z), \quad i = 1, 2.
\label{CLR}
\end{eqnarray}

With Eq. (\ref{CLR}), the decay width for $ l_j \to l_i + \gamma $ can be expressed as\cite{fzb}

\begin{eqnarray}
\Gamma(l_j \to l_i + \gamma) = \frac{e^2}{16\pi} m_j^5 ( |C_2^L|^2 + |C_2^R|^2 ).
\end{eqnarray}
\section{Numerical analysis}
\subsection{Numerical analysis for lepton flavor violating decays}
In this section, we perform numerical calculations of LFV processes within the framework of the $U(1)_X$VLFM. The fixed parameters include
\begin{eqnarray}
&&Q_a=Q_b=1,~~Y_{XN1}=Y_{XN2}=0,~~Y_{XN3}=0.5,~~Y_{PN}=1.0,\nonumber\\&&
Y_{\nu 11} = -3.4717 \times 10^{-5},~~Y_{\nu 12} = -1.8792 \times 10^{-4},
~~Y_{\nu 13} = 1.7292 \times 10^{-4},\nonumber\\&&
Y_{\nu 22} = 1.4 \times 10^{-5},~~Y_{\nu 23} = -1.91 \times 10^{-4},~~Y_{\nu 33} = -2.197 \times 10^{-5},
\nonumber\\&&Y_{\nu ij}=Y_{\nu ji},~\texttt{for} ~ i \neq j.
\end{eqnarray}

and the charged lepton masses $m_e$, $m_\mu$, $m_\tau$ take their experimental values. The neutrino Yukawa couplings $Y_{\nu ij}$ are chosen such that the resulting light neutrino mass squared differences and mixing angles are consistent with current oscillation data. {The loop-level contributions involving
neutrinos are strongly suppressed by the corresponding
Yukawa couplings and thus have a negligible impact on our results. In our numerical analysis, we therefore adopt the Yukawa coupling parameters obtained for the normal mass ordering (NO). The numerical values for the normal ordering are given below.
\begin{eqnarray}
&&\hspace{0cm}\Delta m^2_\odot=0.0000753~{\rm eV^2},~~~\Delta m^2_A=0.002529~{\rm eV^2},
\nonumber\\&&\hspace{0cm}\sin^2\theta_{13}=0.022,~~~\sin^2\theta_{23}=0.553,
~~~\sin^2\theta_{12}=0.307. \label{zwz}
\end{eqnarray}}
Through parameter scans, we take the current experimental upper limits at $90\%$ confidence level as criteria, and present numerical results for the three processes $\mu \to e\gamma$, $\tau \to e\gamma$, and $\tau \to \mu\gamma$ with illustrative figures.
\subsubsection{$\mu \to e\gamma$}
\begin{figure}[ht]
\setlength{\unitlength}{5mm}
\centering
\includegraphics[width=4.5in]{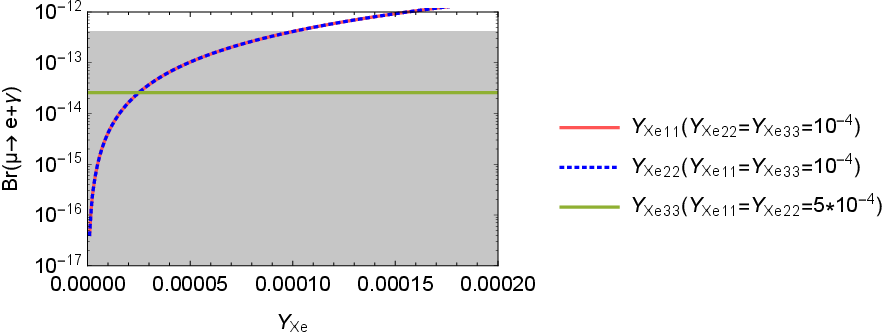}
\setlength{\unitlength}{5mm}
\caption{$Br(\mu \to e\gamma)$ varies with the coupling parameter $Y_{Xe}$, {with $v=246~\rm{GeV}$,~~$v_S=1900~\rm{GeV}$,~~$v_P=4200~\rm{GeV}$,~~$g_{YX}=0.15$,~~$g_X=0.6$}. The gray shaded area represents the  $90\%$ confidence level upper limit given by the MEG experiment, $Br(\mu \to e\gamma) < 4.2\times10^{-13}$.}{\label {AT1}}
\end{figure}
In Fig.~\ref{AT1}, we present the impact of the Yukawa coupling parameter $Y_{Xe}$ on the branching ratio $\mathrm{Br}(\mu \to e\gamma)$ of the LFV process $\mu \to e\gamma$. The horizontal axis denotes the diagonal entries $Y_{Xe11}$, $Y_{Xe22}$, and $Y_{Xe33}$ of the Yukawa matrix $Y_{Xe}$. When scanning over one matrix element, the other two are set to fixed values.

The results show that $Y_{Xe11}$ or $Y_{Xe22}$ dominates the process, with the branching ratio growing obviously as the coupling parameter increases. The two curves almost coincide, indicating that these two parameters contribute comparably to the $\mu \to e\gamma$ process. When $Y_{Xe11}$ or $Y_{Xe22}$ exceeds approximately {$1\times10^{-4}$}, the predicted branching ratio surpasses the experimental upper limit, thereby setting a clear upper bound on these two parameters. In contrast, when {$Y_{Xe11}=Y_{Xe22}=5\times10^{-4}$} is fixed and $Y_{Xe33}$ is varied, the branching ratio shows very little variation, remaining stable around {$3\times10^{-14}$}, always within the allowed region. This indicates that the contribution from the third-generation Yukawa coupling is negligible. This behavior is consistent with the loop-induced mechanism of LFV.
The $\mu \to e\gamma$ decay is forbidden in the SM. In this model, it is generated through loops including the mixing of new vector-like leptons
with SM leptons. Its branching ratio grows with the increase of the Yukawa coupling. This analysis reveals the selective contribution of different flavor couplings to the $\mu \to e\gamma$ process. It also provides direct constraints for further parameter scans and phenomenological studies.
\begin{figure}[ht]
\setlength{\unitlength}{5mm}
\centering
\includegraphics[width=4.2in]{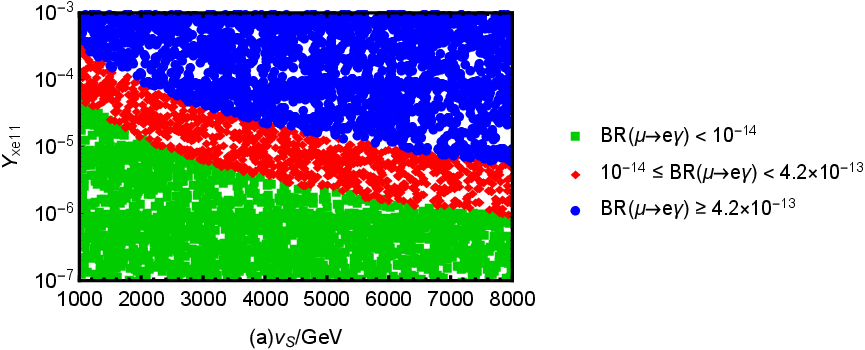}
\setlength{\unitlength}{5mm}
\centering
\includegraphics[width=4in]{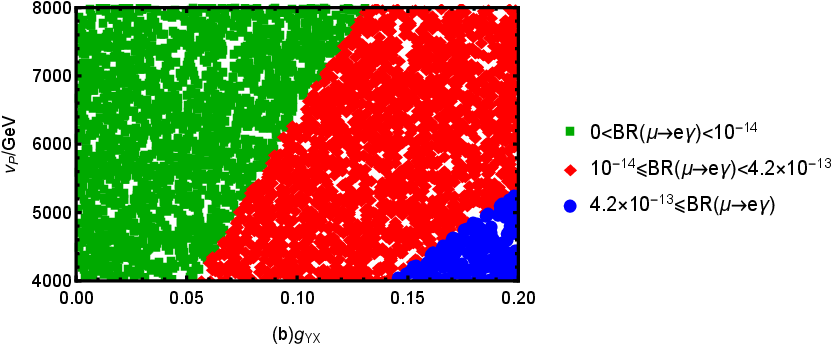}
\setlength{\unitlength}{5mm}
\caption{Parameter space random scan under the constraint of
$\text{Br}(\mu \to e\gamma)$ process.}{\label {AT2}}
\end{figure}

In the $U(1)_X$VLFM, we perform a two-dimensional parameter scan to study the distribution of the $\mu\rightarrow e\gamma$ branching ratio as a function of key parameters, and constrain the parameter space using the MEG experimental upper limit $\mathrm{Br}(\mu\rightarrow e\gamma)<4.2\times10^{-13}$.

Fig.\ref{AT2} (a) shows the parameter scan in the $Y_{Xe11}$ versus {$v_S$} plane: the branching ratio grows with increasing $Y_{Xe11}$ and {$v_S$}. At low {$v_S$} ($\lesssim 2$ TeV), the flavor-violating coupling is constrained to {$Y_{Xe11}\lesssim 10^{-4}$; when {$v_S$} rises to $8$ TeV, the constraint shrinks to $Y_{Xe11}\lesssim 10^{-5}$.
The reason should be that the flavor violating term is ${v_S} Y_{XE}^T$}.

Fig.\ref{AT2} (b) shows the scan in the $g_{YX}$ versus $v_P$ plane. The results indicate that the branching ratio grows linearly with increasing $g_{YX}$ and is suppressed with increasing $v_P$. The green region (${0<\mathrm{Br}<10^{-14}}$) covers almost the entire parameter space with $g_{YX}\lesssim{0.1}$ when $v_P\gtrsim{4}$ TeV; the red region (${10^{-14}\le \mathrm{Br}<4.2\times10^{-13}}$) mainly lies in the diagonal area of the plot; the blue excluded region (${4.2\times10^{-13}\le \mathrm{Br}}$) appears only in the parameter region with low $v_P$ ($\lesssim{5}$ TeV) and large $g_{YX}$ ($\gtrsim{0.15}$), illustrating the combined effect of coupling strength and symmetry-breaking scale on the branching ratio.

\begin{figure}[ht]
\setlength{\unitlength}{5mm}
\centering
\includegraphics[width=3.5in]{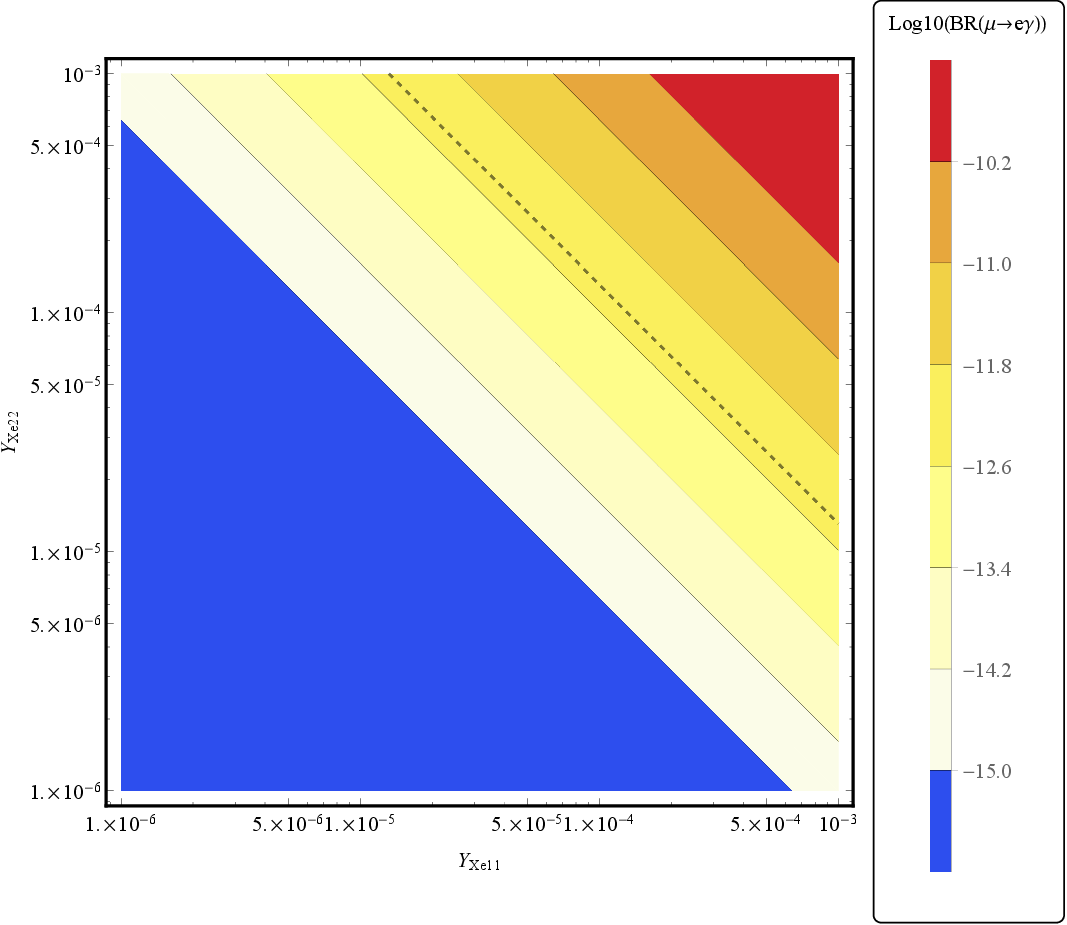}
\setlength{\unitlength}{5mm}
\caption{Contour plot of $\log_{10}\text{Br}(\mu\to e\gamma)$ in the $Y_{Xe11}$-$Y_{Xe22}$ plane, {with $v=246~\rm{GeV}$,~~$v_S=1900~\rm{GeV}$,~~$v_P=4200~\rm{GeV}$,~~$g_{YX}=0.15$,~~$g_X=0.6$}.}{\label {AT3}}
\end{figure}

The contour Fig.~\ref{AT3} shows the distribution of the branching ratio of the LFV decay process $\mu\to e\gamma$ in the $Y_{Xe11}$--$Y_{Xe22}$ plane of Yukawa coupling parameters, with both axes on logarithmic scales and the color representing $\log_{10}\text{Br}(\mu\to e\gamma)$. The results further verify that the branching ratio increases significantly with increasing $Y_{Xe11}$ and $Y_{Xe22}$. The linear distribution of the contour lines is consistent with the loop-induced LFV mechanism. The region in the upper-right of the dashed line corresponds to the parameter space with $\text{Br}(\mu\to e\gamma) > 4.2\times10^{-13}$, which is excluded by the current MEG experimental upper bound; the region in the lower-left of the dashed line satisfies both the current and future MEG II experimental sensitivity requirements, representing the safe parameter region of the model. This result clearly defines the experimentally allowed range for the first- and second-generation lepton Yukawa couplings and reveals the flavor structure characteristics of the model.
\subsubsection{$\tau \to e\gamma$}
\begin{figure}[ht]
\setlength{\unitlength}{5mm}
\centering
\includegraphics[width=4in]{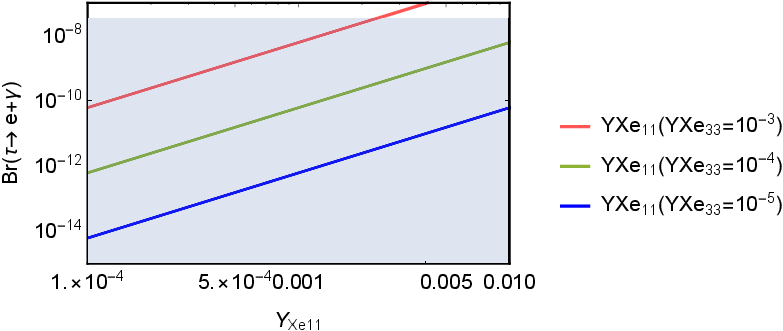}
\setlength{\unitlength}{5mm}
\caption{Branching ratio of $\tau\to e\gamma$ as a function of $Y_{Xe11}$ for different values of $Y_{Xe33}$, {with $v=246~\rm{GeV}$,~~$v_S=1900~\rm{GeV}$,~~$v_P=4200~\rm{GeV}$,~~$g_{YX}=0.15$,~~$g_X=0.6$}.}
\label{BT1}
\end{figure}

The branching ratio of $\tau\to e\gamma$ as a function of the Yukawa coupling $Y_{Xe11}$ in the $U(1)_X{\text{VLFM}}$ is shown in Fig.~\ref{BT1}. During the parameter scan, $Y_{Xe22}$ is fixed to $10^{-4}$, and three different values of $Y_{Xe33}$ ($10^{-3}$, $10^{-4}$, and $10^{-5}$) are considered. Both axes are on logarithmic scales, and the gray shaded region indicates the experimentally allowed range of the branching ratio. All three curves show an increasing trend on the log-log plot. This behavior is consistent with the loop-induced LFV mechanism. As $Y_{Xe33}$ increases from {$10^{-4}$ to $10^{-2}$,} the branching ratio rises by four orders of magnitude, demonstrating a significant synergistic contribution of the third-generation Yukawa coupling to the flavor-violating amplitude in this process. Together with the fixed {$Y_{Xe22}=5\times10^{-3}$}, these results reveal the dependence of the $\tau\to e\gamma$ process on the first- and third-generation Yukawa couplings, complementing the behavior of the $\mu\to e\gamma$ process dominated by the first two generations, thus confirming the flavor structure of the model.
\begin{figure}[ht]
\setlength{\unitlength}{5mm}
\centering
\includegraphics[width=4in]{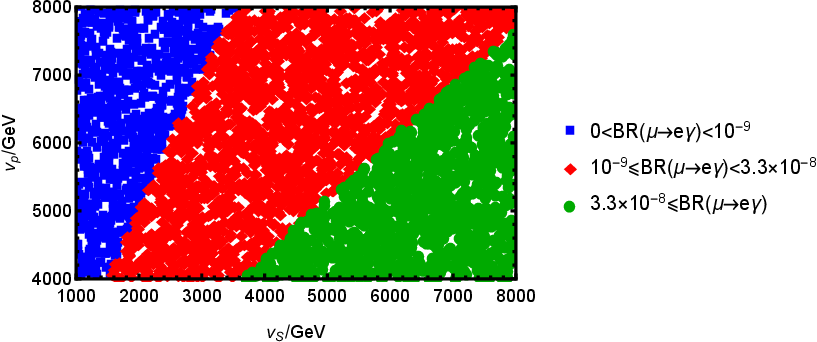}
\setlength{\unitlength}{5mm}
\caption{Distribution of $\text{Br}(\tau\to e\gamma)$ in the {$v_S$}-$v_P$ plane.}{\label {BT2}}
\end{figure}

In the $U(1)_X$VLFM, the distribution of the $\tau\to e\gamma$ branching ratio in the {$v_S$}-$v_P$ plane is shown in Fig.~\ref{BT2}. The result shows that the branching ratio increases significantly as $v_P$ decreases, and increases significantly as {$v_S$} increases, indicating that both jointly determine the strength of the loop contribution. $v_P$, as the new physics mass scale, dominates the suppression effect, while {$v_S$} enhances the flavor-violating contribution through flavor mixing or vertex couplings. The competition between these two effects ultimately leads to the oblique distribution pattern in the figure. This result clearly delineates the allowed parameter space in the {$v_S$}-$v_P$ plane, provides key constraints on the model, and reveals the regulatory role of different VEVs in the LFV process.
\begin{figure}[ht]
\setlength{\unitlength}{5mm}
\centering
\includegraphics[width=3.5in]{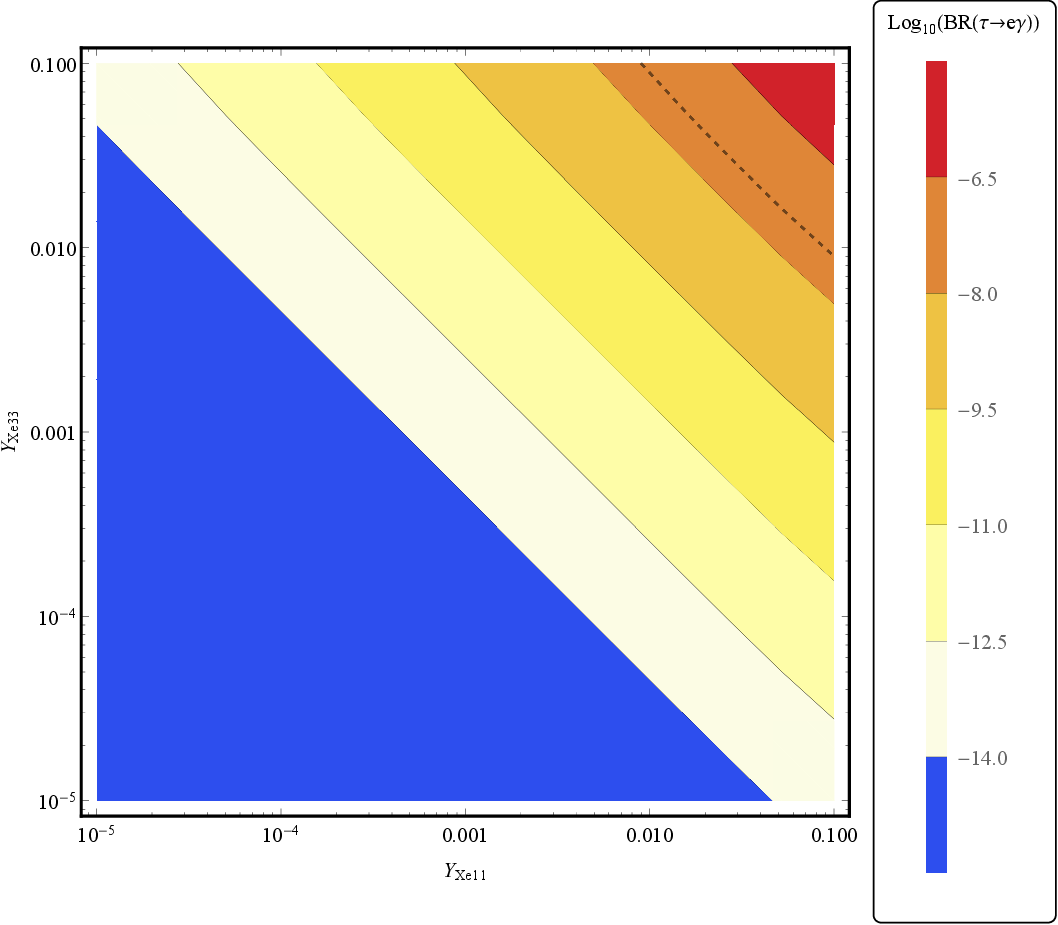}
\setlength{\unitlength}{5mm}
\caption{Contour plot of $\log_{10}\text{Br}(\tau\to e\gamma)$ in the $Y_{Xe11}$-$Y_{Xe33}$ plane, {with $v=246~\rm{GeV}$,~~$v_S=1900~\rm{GeV}$,~~$v_P=4200~\rm{GeV}$,~~$g_{YX}=0.15$,~~$g_X=0.6$}.}{\label {BT3}}
\end{figure}

Fig.~\ref{BT3} shows the distribution of the $\tau\to e\gamma$ branching ratio in the $Y_{Xe11}$--$Y_{Xe33}$ plane of the $U(1)_X{\text{VLFM}}$ , with both axes on logarithmic scales. It can be seen that the branching ratio increases synergistically with $Y_{Xe11}$ and $Y_{Xe33}$, and the contour lines are linear. This confirms that the branching ratio is proportional to the product of the two coupling parameters, consistent with the loop-induced mechanism. The red/orange regions outside the dashed line in the upper right indecates branching ratios above the experimental upper limit $3.3\times10^{-8}$. The region inside the dashed line in the lower left defines the experimentally allowed range for the first- and third-generation lepton Yukawa couplings, reveals the dependence of $\tau\to e\gamma$ on cross-generational couplings, complements the analysis of $\mu\to e\gamma$, and verifies the consistency of the lepton flavor structure in the model.
\subsubsection{$\tau \to \mu\gamma$}
\begin{figure}[ht]
\setlength{\unitlength}{5mm}
\centering
\includegraphics[width=4in]{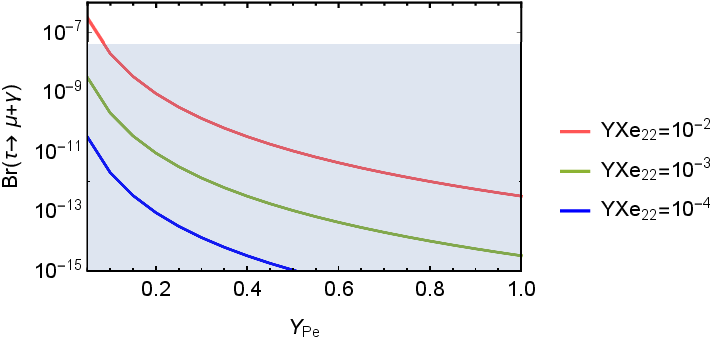}
\setlength{\unitlength}{5mm}
\caption{Branching ratio of $\tau\to\mu\gamma$ as a function of $Y_{Pe}$ for different values of $Y_{Xe22}$, {with $v=246~\rm{GeV}$,~~$v_S=1900~\rm{GeV}$,~~$v_P=4200~\rm{GeV}$,~~$g_{YX}=0.15$,~~$g_X=0.6$}.}{\label {CT1}}
\end{figure}
 In Fig.~\ref{CT1}, we fix $Y_{Xe22}$ to three different values $10^{-2}$, $10^{-3}$ and $10^{-4}$ and plot the branching ratio of $\tau\to\mu\gamma$ as a function of $Y_{Pe}$. The branching ratio decreases monotonically with increasing $Y_{Pe}$ and increases significantly with larger $Y_{Xe22}$. For $Y_{Xe22}=10^{-2}$, the branching ratio exceeds the experimental upper limit at low $Y_{Pe}$. For $Y_{Xe22}=10^{-4}$, the branching ratio stays well below the limit, providing a wide allowed parameter space.
  This result reveals the dependence of $\tau\to\mu\gamma$ on the second-generation coupling $Y_{Xe22}$ and the new coupling $Y_{Pe}$. The result gives experimental constraints on the model parameters, and complements the analyses of $\mu\to e\gamma$ and $\tau\to e\gamma$. It confirms that the consistency of the lepton flavor structure in the model.

\begin{figure}[ht]
\setlength{\unitlength}{5mm}
\centering
\includegraphics[width=4in]{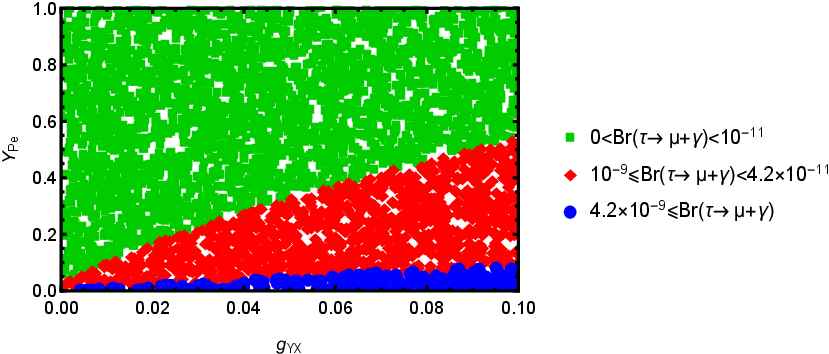}
\setlength{\unitlength}{5mm}
\caption{Parameter space scan in the $g_{YX}$-$Y_{Pe}$ plane under the constraint of $\text{Br}(\tau\to\mu\gamma)$.}{\label {CT2}}
\end{figure}

The distribution of the branching ratio for $\tau\to\mu\gamma$ in the $g_{YX}$--$Y_{Pe}$ plane is presented in the Fig.~\ref{CT2}. The branching ratio increases significantly with $g_{YX}$ and decreases with $Y_{Pe}$. At large $g_{YX}$ and small $Y_{Pe}$, the flavor-violating interaction strength in the loop diagrams is high and the mass scale of new particles is low, leading to a branching ratio above the experimental upper limit (blue region). As $g_{YX}$ decreases or $Y_{Pe}$ increases, the flavor-violating loop contributions are suppressed, and the branching ratio falls into the experimentally allowed range (red region), and is strongly suppressed in the low $g_{YX}$, high $Y_{Pe}$ region (green region). This behavior is fully consistent with the loop-induced LFV mechanism in the model: $g_{YX}$ directly controls the interaction strength of the flavor-violating vertex, while $Y_{Pe}$ suppresses the branching ratio by affecting the mass scale of new scalar particles. This result clearly defines the allowed parameter regions in the $g_{YX}$--$Y_{Pe}$ plane under current experimental constraints.
\begin{figure}[ht]
\setlength{\unitlength}{5mm}
\centering
\includegraphics[width=3.5in]{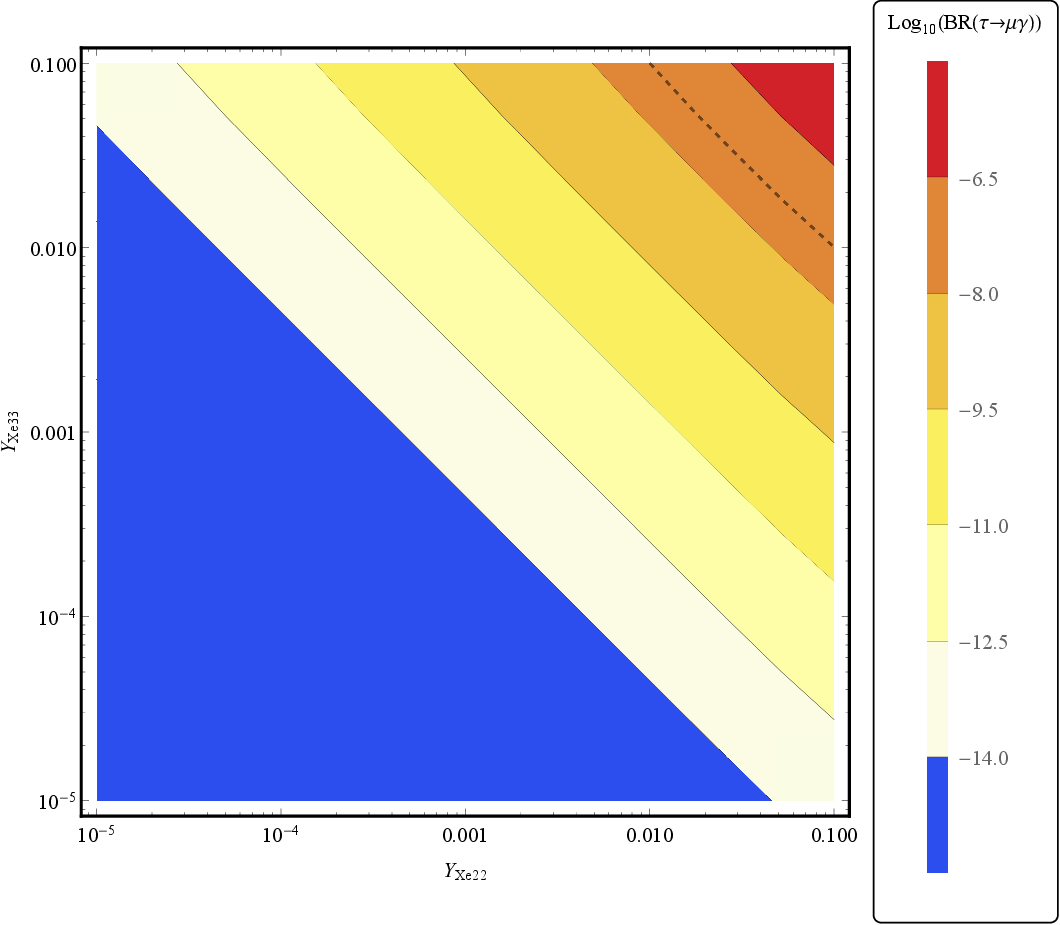}
\setlength{\unitlength}{5mm}
\caption{Contour plot of $\log_{10}\text{Br}(\tau\to\mu\gamma)$ in the $Y_{Xe22}$-$Y_{Xe33}$ plane, {with $v=246~\rm{GeV}$,~~$v_S=1900~\rm{GeV}$,~~$v_P=4200~\rm{GeV}$,~~$g_{YX}=0.15$,~~$g_X=0.6$}.}{\label {CT3}}
\end{figure}

The contour plot shows the distribution of the $\tau\to\mu\gamma$ branching ratio in the $Y_{Xe22}$--$Y_{Xe33}$ plane on logarithmic scales. As shown in Fig.~\ref{CT3}, the branching ratio increases synergistically with $Y_{Xe22}$ and $Y_{Xe33}$, which is consistent with the loop-induced mechanism. The region outside the dashed line in the upper right has a branching ratio above the experimental upper limit $4.2\times10^{-8}$ (excluded).
This result shows the experimentally allowed range for the second- and third-generation lepton Yukawa couplings. It reveals the dependence of $\tau\to\mu\gamma$ on cross-generation couplings and complements the analyses of $\mu\to e\gamma$ and $\tau\to e\gamma$.
{\subsection{Constraints from LFV Higgs decays}}
We analyze the tree-level lepton-flavor-violating Higgs decay process \(h\to l_i l_j\). In the lepton-flavor-violating Higgs decay $h\to l_i l_j$, charged-lepton loop diagrams are suppressed by loop factors, while the amplitudes involving neutrinos receive double suppression from both loop integrals and tiny neutrino Yukawa couplings. Accordingly, the tree-level contribution dominates.
The corresponding effective amplitude can be written as
{\begin{eqnarray}
&& \mathcal{M} = \bar{l}_i \big(H_L P_L + H_R P_R\big) l_j \, h,
\end{eqnarray}}
$H_{L,R}^{hl_il_j}$ denote the left- and right-handed couplings in the $h$-$l_i$-$l_j$ interaction vertices. The specific forms have already been given in Eq.(\ref{hll}).

Then, the branching ratio of $h \to l_i^{\pm} l_j^{\mp}$ is defined as
{\begin{eqnarray}
&& Br\left(h \to l_i^{\pm} l_j^{\mp}\right) = \frac{1}{16\pi}\frac{m_h}{\Gamma_h}\left(\left|H_L\right|^2 + \left|H_R\right|^2\right),
\end{eqnarray}}
here $\Gamma_h \simeq \Gamma_h^{SM} \simeq 4.1\times 10^{-3}\,\text{GeV}$.
$\Gamma_h$ represents the total decay width of the Higgs boson.
Some parameters are set as below, and the rest are adopted from the original work
{\begin{eqnarray}
&&\lambda_{H} = -0.14,\quad \lambda_{P} = -0.05,\quad \lambda_{S} = -0.05,
\nonumber\\
&&\lambda_{HP} = -0.01,\quad \lambda_{HX} = -0.05,\quad \lambda_{PX} = -0.05,
\end{eqnarray}}
The vacuum stability conditions
{\begin{eqnarray}
&&\mu_{H}^2 = \lambda_{H}\,v^2 + \frac{1}{2}\lambda_{HP}\,v_P^2 + \frac{1}{2}\lambda_{HX}\,v_S^2,
\nonumber\\
&&\mu_{X}^2 = \lambda_{X}\,v_S^2 + \frac{1}{2}\lambda_{HX}\,v^2 + \frac{1}{2}\lambda_{PX}\,v_P^2,
\nonumber\\
&&\mu_{P}^2 = \lambda_{P}\,v_P^2 + \frac{1}{2}\lambda_{HP}\,v^2 + \frac{1}{2}\lambda_{PX}\,v_S^2.
\end{eqnarray}}

 We calculate the branching ratios of the process $h\to l_i l_j$ at tree level as follows
\begin{center}
$\text{Br}(h\to e\mu)$ $\simeq$ $10^{-18}\sim 10^{-21}$,

$\text{Br}(h\to e\tau)$ $\simeq$ $10^{-16}\sim 10^{-19}$,

$\text{Br}(h\to \mu\tau)$ $\simeq$ $10^{-16}\sim 10^{-19}$,
\end{center}
and the results are far below the experimental upper limits\cite{PDG}
\begin{center}
$\text{Br}(h\to e\mu) < 4.4 \times 10^{-5}$,

$\text{Br}(h\to e\tau) < 2.0 \times 10^{-3}$,

$\text{Br}(h\to \mu\tau) < 1.5 \times 10^{-3}$.
\end{center}
Throughout the parameter region allowed by the $l_j\to l_i\gamma$ constraints, the predicted branching ratios of $h\to l_i l_j$ remain well below the current experimental upper limits.
{\subsection{Constraints from $Z'$ searches}}
The mass of $Z'$ boson is predicted $m_{Z'} > 5.1 \rm{TeV}$ based on the search for dilepton resonances at ATLAS and CMS\cite{ATLAS,Z1,Z2}.

The mass of the neutral gauge boson $Z'$ is obtained by diagonalizing the corresponding gauge boson mass matrix.
The resulting mass eigenvalue is given by Eq.(\ref{zp}) and simplified as
\begin{eqnarray}
&&m^2_{Z'}\simeq g_X^2\xi^2,
\end{eqnarray}
where $\xi^2=4(Q_a+Q_b)^2v_P^2+4Q_a^2v_S^2$ and $\xi^2\gg v^2$.

The electroweak gauge couplings are determined from
\begin{eqnarray}
g_1=\frac{e}{c_W},
~~g_2=\frac{e}{s_W},
~~e=\sqrt{4\pi\alpha},
~~s_W=\sqrt{1-\frac{m_W^2}{m_Z^2}},
~~c_W=\frac{m_W}{m_Z}.
\end{eqnarray}
 With these input parameters, we obtain
$m_{Z'}\simeq5.17~\mathrm{TeV}$, satisfying the current experimental constraints.
\section{Conclusion}
In the SM, the theoretical branching ratio of the LFV processes $l_j \to l_i\gamma$ is extremely suppressed by the GIM mechanism. For instance, the predicted value of $\mu \to e\gamma$ is only approximately $10^{-55}$. It is far lower than the current experimental sensitivity ($4.2 \times10^{-13}$). Therefore, once such decays are observed in future experiments, they will become conclusive evidence of new physics beyond the SM. In this paper, under the framework of the $U(1)_X$VLFM, three LFV processes, namely $\mu \to e\gamma$, $\tau \to e\gamma$, and $\tau \to \mu\gamma$, are systematically studied. This model extends the SM group to $SU(3)_C \otimes SU(2)_L \otimes U(1)_Y \otimes U(1)_X$, introducing three generations of right-handed neutrinos, two scalar singlets, and one generation of vector-like quarks, vector-like leptons and vector-like neutrinos. Neutrino mass is generated through the seesaw mechanism and provides an additional contribution to the destruction of lepton flavor. We calculate the relevant Feynman amplitudes and conduct a large-scale parameter scan under the current experimental constraints. This allows us to identify the key sensitive parameters that affect the branching ratio.

The numerical results show that the Yukawa couplings $(Y_ {Xe11} $, $Y_ {Xe22}$, $Y_ {Xe33}$, $Y_ {Pe})$, the gauge mixing parameters ($g_ {YX}$, $g_ {X})$, and the vacuum expectation values ({$v_S$} , $v_P$) significantly affect the branching ratios. Furthermore, the contributions from different generations of couplings show clear selectivity among different processes. The overall rule is as follows: the branching ratio increases with the relevant Yukawa coupling, and decreases as the scalar VEV increases. This is consistent with the loop-induced LFV mechanism. The VEV determines the mass scale of new particles. The heavier the mass, the stronger the suppression.

The $U(1)_X$VLFM introduces a rich source of LFV through the mixing of vector-like leptons with SM leptons. Moreover, additional scalar fields and gauge mixing can make LFV effects stronger. It can generate a sufficiently large branching ratio while meeting the existing experimental constraints, with the branching ratios in some parameter space approaching or even exceeding the current upper limit. This indicates that the model has the potential to be tested or excluded in the next generation of LFV experiments. This study define the experimental allowable intervals of key parameters, reveal the selective contributions of different generations of coupling to each process, and provide a clear theoretical basis and parameter guidance for exploring new physical phenomena beyond the SM.

\begin{acknowledgments}

This work is supported by National Natural Science Foundation of China (NNSFC)
(No.12075074), Natural Science Foundation of Hebei Province
(A2023201040, A2022201022, A2022201017, A2023201041), Natural Science Foundation of
Hebei Education Department (QN2022173), Post-graduate's Innovation
Fund Project of Hebei University (HBU2024SS042), the Project of the China
Scholarship Council (CSC) No. 202408130113. This work is also supported by Funda\c{c}\~{a}o para a Ci\^{e}ncia e a Tecnologia (FCT, Portugal) through the project UID/00777/2025 (https://doi.org/10.54499/UID/00777/2025)

\end{acknowledgments}

\end{document}